# Low-Frequency Electronic Noise in Exfoliated Quasi-1D TaSe$_3$ van Der Waals Nanowires

Guanxiong Liu[1], Sergey Rumyantsev[2,3], Matthew A. Bloodgood[4], Tina T. Salguero[4], Michael Shur[2] and Alexander A. Balandin[1]

[1]Nano-Device Laboratory (NDL) and Phonon Optimized Engineered Materials (POEM) Center, Department of Electrical and Computer Engineering, University of California – Riverside, Riverside, California 92521 USA

[2]Department of Electrical, Computer, and Systems Engineering, Center for Integrated Electronics, Rensselaer Polytechnic Institute, Troy, New York 12180 USA

[3]Ioffe Physical-Technical Institute, St. Petersburg 194021 Russia

[4]Department of Chemistry, University of Georgia, Athens, Georgia 30602 USA

**Abstract**

We report results of investigation of the low-frequency electronic excess noise in quasi-1D nanowires of TaSe$_3$ capped with quasi-2D *h*-BN layers. Semi-metallic TaSe$_3$ is a quasi-1D van der Waals material with exceptionally high breakdown current density. It was found that TaSe$_3$ nanowires have lower levels of the normalized noise spectral density, $S_I/I^2$, compared to carbon nanotubes and graphene (*I* is the current). The temperature-dependent measurements revealed that the low-frequency electronic 1/*f* noise becomes the 1/*f*$^2$-type as temperature increases to ~400 K, suggesting the onset of electromigration (*f* is the frequency). Using the Dutta–Horn random fluctuation model of the electronic noise in metals we determined that the noise activation energy for quasi-1D TaSe$_3$ nanowires is approximately $E_P \approx 1.0$ eV. In the framework of the empirical noise model for metallic interconnects, the extracted activation energy, related to electromigration, is $E_A = 0.88$ eV, consistent with that for Cu and Al interconnects. Our results shed light on the physical mechanism of low-frequency 1/*f* noise in quasi-1D van der Waals semi-metals and suggest that such material systems have potential for ultimately downscaled local interconnect applications.

The investigations of the two-dimensional layers and heterostructures revealed new physics and demonstrated promising applications [1-14]. Starting with graphene [3-5], and spreading to a wide range of layered van der Waals materials [6-10], successful isolation of individual atomic layers from their respective bulk crystals by mechanical exfoliation led to the fast growing research activities in the 2D materials. In contrast to the layered van der Waals materials that yield 2D crystals, materials, such as $TaSe_3$ and $TiS_3$ [15-17] yield the quasi-one-dimensional (1D) van der Waals crystal structures. These materials belong to the group of the transition metal trichalcogenides $MX_3$ (where M = Mo, W, and other transition metals; X = S, Se, Te). In the monoclinic crystal structure of $TaSe_3$, the trigonal prismatic $TaSe_3$ units form continuous chains extending along the *b* axis and leading to fiber- and needle-like crystals with anisotropic semi-metallic or metallic properties. The quasi-1D atomic threads are weakly bound in bundles by the van der Waals forces. As a consequence, the mechanical exfoliation of the $MX_3$ crystals results not in the 2D layers but rather in the quasi-1D van der Waals nanowires. We have recently demonstrated quasi-1D $TaSe_3$ nanowires with the record high current density exceeding $J_B$~10 $MA/cm^2$, which is an order of magnitude larger than that for the Cu interconnects [18].

In this Letter, we report on the excess low-frequency electronic noise in quasi-1D nanowires of $TaSe_3$ capped with the quasi-2D *h*-BN layers. The *h*-BN capping was used as a surface passivation protecting from environmental exposure. The measurements were performed in the temperature range from 300 K to 430 K in order to elucidate the physical mechanism of the low-frequency noise and determine the noise activation energies. The low-frequency noise is a ubiquitous phenomenon present in all kinds of electronic devices including field-effect transistors, diodes and interconnects [19-21]. The frequency dependence and temperature characteristics of the spectral density of the low-frequency noise have been used as reliability metrics for interconnects in Si complementary metal-oxide-semiconductor (CMOS) technology. The low-frequency noise data at elevated temperatures are directly correlated with the electromigration failure mechanisms in interconnects [22-24]. It was suggested that the low-frequency noise at high temperatures originates from the vacancy migration along the grain boundaries [22, 25]. Hence, the noise can reveal information on the electromigration failure of interconnects and metal thin films. The noise activation energies extracted from the two commonly accepted physical models, the Dutta–Horn model [21, 24] and the empirical noise model in metals [22, 26, 27], have shown an excellent agreement with the activation energies obtained from the industry standard electromigration mean-time-to-failure (MTF) tests. These considerations explain additional practical motivations for investigation of the low-frequency noise in quasi-1D nanowires of TaSe$_3$, which exhibit promise as ultimately downscaled local interconnects owing to their single-crystal atomic thread structure and exceptionally high breakdown current density [18].

The bulk TaSe$_3$ crystals were synthesized by the chemical vapor transport (CVT) method and then were solvent exfoliated (see Methods). The high quality of both bulk and exfoliated quasi-1D TaSe$_3$ samples has been confirmed using a variety of experimental techniques including high-resolution transmission electron microscopy, Raman spectroscopy, energy dispersive spectroscopy, powder X-ray diffraction, and electron probe micro analysis, as reported previously [18]. The devices for the low-frequency noise measurement were fabricated by the mechanical exfoliation of the quasi-1D TaSe$_3$ nanowires on Si/SiO$_2$ substrates. We used the dry transferred *h*-BN flakes to protect the quasi-1D TaSe$_3$ nanowire channels from oxidation and chemical exposure during the fabrication processes. The devices were fabricated by standard e-beam lithography and e-beam metal evaporation [28]. In order to make an electrical connection to the quasi-1D channels we opened the contact area by dry etching the *h*-BN thin film and the underlying TaSe$_3$ nanowire while leaving the edge of quasi-1D channels in contact with the metal. Figure 1 (a) shows a high-resolution transmission electron microscopy (TEM) image of TaSe$_3$ crystal prepared by CVT method together with a schematic of the atomic plane arrangements. A scanning electron microscopy (SEM) of the long TaSe$_3$ nanowires is presented in Figure 1 (b). Figure 1 (c) is a SEM image of representative devices used for the current voltage (I-V) and electronic noise measurements.

[Figure 1 (a-c)]

Figure 2 presents typical I-V characteristics of the devices with different channel lengths. The perfectly linear I-Vs at low electric fields indicate a high-quality electrical contact between the metal electrode and the quasi-1D TaSe$_3$ nanowire. The *h*-BN capping was essential for achieving a good Ohmic contact between the metal electrodes and the TaSe$_3$ channel. The inset to Figure 2 shows an optical image of a representative device used in this study. The light blue stripe between metal electrodes is the quasi-1D TaSe$_3$ nanowire while the green region is the *h*-BN capping layer. The red background is the Si/SiO$_2$ substrate.

[Figure 2]

The low-frequency noise measurements were performed both at UCR and RPI using in-house built experimental setups. The equipment included a "quiet" battery, a potentiometer biasing circuit, a low noise amplifier and a spectrum analyzer. The details of our noise measurement procedures have been reported elsewhere [28 - 29]. Figure 3 (a) shows a typical normalized noise spectral density $S_I/I^2 \propto 1/f^\gamma$ with $\gamma = 1.16$ (here *I* is the channel source – drain current) in the TaSe$_3$ nanowires with the 20 nm × 110 nm cross-section. The measurements were performed at room temperature (RT). The noise spectral density was determined for the currents in the 1 µA – 5 µA range, which is sufficiently small to exclude the current induced effects such as electromigration. The latter is confirmed by the perfectly quadratic dependence of the noise spectral density on the source – drain current, i.e. $S_I \propto I^2$. The noise spectral density as a function of the source-drain current is shown in the inset to Figure 3 (a). The noise spectrum was collected for the two-terminal device. Since the contact resistance is very small compare to that of the channel, the contact noise was negligible.

The first interesting observation from the experimental data is that the noise level in TaSe$_3$ nanowires is rather low in comparison with other low-dimensional materials, such as carbon nanotubes and graphene, which also have been proposed as the interconnect applications. The normalized noise spectral density, $S_I/I^2$, for TaSe$_3$ nanowires as a function of the resistance is shown in Figure 3 (b). The data points for carbon nanotubes [30-32] and graphene [29, 33] are presented for comparison. The guide line is the noise amplitude, $A = S_I/I^2 \times f$, which was empirically obtained for the carbon nanotubes [30]. The devices with a lower resistance tend to have a lower level of the low-frequency noise. As one can see, the noise level, expressed as $S_I/I^2$, in quasi-1D TaSe$_3$ nanowires is about one order of magnitude lower than that in graphene [29, 33]. It is comparable to the lowest reported values for the carbon nanotubes [30-32]. Most of the tested carbon nanotube devices had a larger resistance and, as a result, a larger noise spectral density. The higher levels of 1/*f* noise typically are associated with a higher density of structural defects, as well as mechanical and electromigration damage [22, 25].

[Figure 3 (a-b)]

The resistivity of quasi-1D TaSe$_3$ nanowires increases with temperature, similar to other metallic materials. The inset to Figure 4 shows the resistivity change in the temperature range from 298 K to 450 K. The gradual and relatively slow resistance increase below 420 K is due to increasing electron–phonon scattering, which is common in metals. The sharp increase in resistance at the temperatures above 420 K is likely related to the onset of electromigration as commonly observed in the conventional interconnect reliability tests [24, 34]. The TaSe$_3$ nanowire noise spectral density and the frequency exponent $\gamma$ rapidly grows with temperature (see Figure 4). The noise level at 378 K increases by a factor of ×10 at 100 Hz and a factor of ×300 at 1 Hz compared to that at 298 K. The noise magnitude grows faster at the lower frequencies, changing the type of the frequency dependence of the normalized noise spectral density, $S_I/I^2$, at high temperature.

[Figure 4]

To better understand the evolution of the low-frequency noise with temperature, we plotted the frequency exponent $\gamma$ and the normalized noise spectral density, $S_I/I^2$, as the functions of temperature in the range from 298 K to 431 K. The experimental data points in Figure 5 (a-b) are shown for two representative devices with different channel cross-section areas at the frequency $f=10$ Hz. The frequency exponent $\gamma$ increases from ~1.16-1.24 at 298 K to 1.72-1.75 at 350 K. At temperatures above 350 K, $\gamma$ remains nearly constant at 1.71-1.75 (see Figure 5 (a)). In Figure 5 (b), one can see that the noise level increases by two orders of magnitude as temperature changes from 298 K to 400 K. The data in Figure 5 (b) suggest that the noise spectral density attains its maximum value at T=400 K and then reveal a saturation and a slowly decreasing trend. This behavior is analogous to that in thin films of Ag and Cu [35, 36]. In the region of the fast growth, $S_I/I^2 \sim T^s$ with $s=21$ for the shown data set for two representative devices (see Figure 5 (b)).

[Figure 5]

We now turn to the analysis of the experimental data using two different physical models. The Dutta–Horn model describes the temperature dependent low-frequency noise by the thermally activated random fluctuations [35]. Within this model, the noise spectral density, $S(f,T)$, can be approximated as

$$S(f,T) \propto \frac{kT}{2\pi f} D(E), \qquad (1)$$

where $D(E)$ is the distribution density function of the activation energies $E$ and $S(f, T)$ is the spectral noise density. In the limit of the constant $D(E)$ the model leads to the exact 1/$f$ noise ($\gamma$=1). The experimentally observed deviations from $\gamma$=1 and temperature dependence of the noise indicate that $D(E)$ is not a constant. One should note here that the maximum in the $S(f, T)$ dependence on temperature does not always coincide with the maximum of $D(E)$ function. The temperature dependence of the frequency exponent $\gamma$ is expressed as [35]

$$\gamma(T) = 1 - \frac{1}{\ln(2\pi f \tau_0)} \left( \frac{\partial \ln S(f,T)}{\partial \ln T} - 1 \right). \qquad (2)$$

Here $\tau_0$ is characteristic time constant referred to as the attempt escape time. In the original Dutta–Horn model, the time constant $\tau_0$ was assumed to be ~$10^{-14}$ s which is the typical attempt escape time for random processes in solids. The exact value of $\tau_0$ does not strongly affect the low-frequency noise analysis. The blue curve in Figure 5 (a) is a polynomial fit to the experimental data used in the further analysis.

In Figure 5 (b), the blue line is calculated using Eq. (2) with the approximated dependence of the frequency exponent $\gamma$ on temperature shown in Figure 5 (a). One can see that the Dutta-Horn model describes accurately the temperature dependence of the noise for T < 380 K. The saturation or even the decrease of noise at higher temperatures cannot be explained by this model. The latter is likely related to the onset of another low-frequency noise mechanism. For this reason, applying of the Dutta–Horn model for temperatures above ~400 K has to be done with caution. The activation energy $E_P$, at which the distribution function $D(E)$ attains its peak is related to temperature as

$$E_P = -kT \ln(2\pi f \tau_0). \qquad (3)$$

The activation energy distribution calculated from the noise data in metals usually has a peak at $E_P \approx 1$ eV [35]. Using the data for a representative device heated all the way to 430 K we determined that the energy distribution reaches its maximum at $E_P \approx 1.0$ eV for TaSe$_3$ nanowires as well (data set for "Device 1" in Figure 5 (b)). In a more conservative approach, increasing the temperature only to ~390 K, within the application domain of the Dutta–Horn model (data set for "Device 2" in Figure 5 (b)), we determine the low bound of the position of the maximum of the activation energy distribution function $D(E)$. This value is somewhat smaller but still close to ~1 eV.

The noise spectral density $S_I/I^2 \sim 1/f^\gamma$ with $\gamma \approx 2$ in metals is often attributed to the effects of electromigration. Although the detailed mechanism of the resistance fluctuations in metal films has not been established yet, many experiment indicate that the low-frequency noise, electromigration, and degradation in metals are related. The $1/f^\gamma$ noise spectral density with $\gamma \geq 2$, which appears at high temperatures, has been widely used to extract the electromigration activation energy for interconnects. The frequency exponents as high as $\gamma > 4$ have been reported [38]. Attributed to the vacancy migration, $1/f^\gamma$ noise has been proposed as a useful diagnostic tool [25]. It was found that the low-frequency noise at high temperature follows the empirical equation [22]

$$S(f) = \frac{Aj^3}{Tf^\gamma} \exp\left(-\frac{E_A}{kT}\right), \qquad (4)$$

where $A$ is a constant, $j$ is the biasing current density and $E_A$ is the activation energy related to electromigration. By measuring the temperature dependence of the noise spectrum one can extract the activation energy, $E_A$, from the Arrhenius plot of $T \times S(f)$. A large number of experiments have demonstrated that the $E_A$ values extracted from the low-frequency noise measurements are very close to those obtained from the conventional mean-time-to-failure test for interconnects fabricated with different technology and various metals, including Al, Al alloys and Cu [21, 23, 24, 37]. The noise measurements have specific advantages over the conventional reliability tests, which require time-dependent resistance measurements of interconnects under high temperature and high current stress for long periods of time (typically from 10 to 100 hours for each set). The noise measurements can be conducted in much shorter period of time. They are also be non-destructive. It was found that the noise level often increases rapidly before the sharp resistance increase, which was defined as the interconnect failure [24].

Figure 6 shows the Arrhenius plot of $T \times S_I/I^2$ vs. $1000/T$ for $f$=10 Hz. The extracted $E_A$ of 0.88 eV is within the range reported for other metal films. Table I summarizes the electromigration activation energy estimated using the 1/$f$ noise measurements for different interconnects. The values of the electromigration activation energy for the Cu and Al interconnects determined from the 1/$f$ noise measurements are 0.76-1.10 eV and 0.67- 1.14 eV, respectively [21, 22, 24, 38, 39]. Our data suggest that quasi-1D TaSe$_3$ nanowires have similar $E_A$ values. Considering the 10× fold higher current carrying capacity and a possibility of ultimate downscaling of the quasi-1D TaSe$_3$ van der Waals material, one can consider TaSe$_3$ to be a promising material for the nanoscale interconnects.

[Figure 6]

In conclusion, we investigated the low-frequency excess noise in quasi-1D nanowires of TaSe$_3$ capped with quasi-2D $h$-BN layers. The TaSe$_3$ nanowires have relatively low level of the 1/$f$ noise in comparison with carbon nanotubes and other low-dimensional materials. The temperature dependent measurements revealed that the electronic excess noise becomes the $1/f^\gamma$ – type with $\gamma$>1 as temperature increases. Using the Dutta–Horn random fluctuation model of 1/$f$ noise in metals and empirical model for metallic interconnects we extracted the noise activation energy for quasi-1D TaSe$_3$ nanowires. Our results indicate that quasi-1D van der Waals materials have potential for the ultimately downscaled local interconnect applications.

**METHODS**

**Material Preparation:** TaSe$_3$ was prepared from elemental tantalum (12.0 mmol, STREM 99.98% purity) and selenium (35.9 mmol, STREM 99.99% purity) with iodine (~6.45 mg/cm$^3$, J.T. Baker, 99.9% purity) as the transport agent. The tantalum/selenium mixture was ground and placed in a 17.78 x 1 cm fused quartz ampule (cleaned in concentrated nitric acid followed by annealing for 12 h at 900 °C). The charged reaction ampule was evacuated and backfilled with Ar 3x while submerged in an acetonitrile/dry ice bath. The flame-sealed, ampule was placed in a three-zone, horizontal tube furnace heated at 20 °C min$^{-1}$ to the final temperature gradient of 700 °C (hot zone) – 680 °C (cool zone). The ampule was held at these temperatures for two weeks before cooling to room temperature. TaSe$_3$ crystals were removed carefully from the quartz ampule, and residual I$_2$ was removed by vacuum sublimation at 50 °C. The isolated yield of silver-black crystals was 90.8%. TaSe$_3$ was chemically exfoliated by sonicating 6 mg of powdered TaSe$_3$ crystals in 10 mL ethanol for 4 h, resulting in a brownish-black suspension. This suspension was then centrifuged at 2600 rpm for 15 min, leaving well-dispersed TaSe$_3$ wires 30 to 80 nm wide in the supernatant.

## Acknowledgements

This project was supported, in part, by the by the Emerging Frontiers of Research Initiative (EFRI) 2-DARE project: Novel Switching Phenomena in Atomic MX$_2$ Heterostructures for Multifunctional Applications (NSF EFRI-1433395). The nano-device fabrication and characterization work at UC Riverside was also supported, in part, by the Semiconductor Research Corporation (SRC) and Defense Advanced Research Project Agency (DARPA) through STARnet Center for Function Accelerated nanoMaterial Engineering (FAME). The quasi-1D TaSe$_3$ devices were fabricated in the UCR Center for Nanoscale Science and Engineering (CNSE). The authors acknowledge useful discussions on 1/$f$ noise with Dr. Michael Levinshtein (Ioffe Institute, St. Petersburg, Russia).

## Author Contributions

A.A.B. conceived the idea, coordinated the project, and contributed to experimental data analysis; T.T.S. supervised material synthesis and contributed to materials analysis; G.L. fabricated and tested devices, analyzed experimental data, and conducted part of the low-frequency noise measurements; M.A.B. synthesized TaSe$_3$ and conducted materials characterization; S.L.R. conducted noise measurements and analyzed experimental data; M.S.S. and M. L. contributed to data analysis. All authors contributed to writing of the manuscript.

**Supplementary Information:**

Details concerning device fabrication are available on the journal web-site for free-of-charge.

**TABLE I:** $E_A$ and $E_p$ Extracted from 1/$f$ Noise Measurements for Different Technologies

| Technology | $E_p$ (eV) - D-H model | $E_A$ (eV) Empirical Model |
|---|---|---|
| Cu direct etch | 1.10 [24] | |
| Cu with TaNTa barrier | 0.76-0.78 [24] | 0.79 [38] |
| Al | 0.69 [21] | |
| Al – Si (1%) | 0.80 [21] | |
| Al – Cu (4%) | 0.89 [21] | |
| Al thin film | | 1.14 [22] |
| Al – Si (0.8%) passivated | | 0.74 [39] |
| Al – Si (1%) – Cu (0.5%) passivated | | 0.78 [39] |
| TaSe$_3$ | 1.01 [this work] | 0.88 [this work] |

**Figure Captions**

**Figure 1:** (a) High resolution scanning transmission electron microscopy image of exfoliated TaSe$_3$ showing pristine metal trichalcogenide chains that extend along the *b* axis. (b) Scanning electron microscopy image of TaSe$_3$ crystals prepared by CVT method. (c) SEM image of representative quasi-1D TaSe$_3$ devices.

**Figure 2:** Current-voltage characteristics of TaSe$_3$ devices with different channel length. The linear characteristics at low voltage indicates good Ohmic contact of TaSe$_3$ channel with the metal electrodes. The optical microscopy image of the devices is shown in the inset. The quasi-1D TaSe$_3$ channel (light blue) is covered with *h*-BN capping layer (green area), which acts as protection layer against oxidation and environmental exposure. The metal electrodes, in contact with TaSe$_3$ channel at the edges, are fabricated by etching through *h*-BN capping layer.

**Figure 3:** (a) Typical noise spectrum of TaSe$_3$ devices at room temperature. The noise spectrum $S_I$ is flowing $1/f^{\gamma}$ dependence with $\gamma=1.1\sim1.2$. The inset shows the noise level at $f=10$ Hz as the function of the channel (source-drain) current spanning from 0.2 µA to 10 µA. The quadratic dependence of the noise spectrum density $S_I$ on the channel current $I$ indicates that the $1/f$ noise measured at this current level originates from the TaSe$_3$ device itself rather than the current induced effects. (b) Normalized noise spectrum density as a function of the resistance for different low-dimensional material systems - quasi-1D TaSe$_3$ nanowires, graphene and carbon nanotubes. For comparison, the empirical relation $A=10^{-11}R$ for the low-frequency noise versus resistance $R$ derived from the carbon nanotube studies [29, 31-33] is also shown. The noise in quasi-1D TaSe$_3$ nanowires is about one order of magnitude lower than that in graphene.

**Figure 4:** Normalized noise spectra density $S_I/I^2$ measured for the quasi-1D TaSe$_3$ nanowire at different temperatures. The $1/f$ noise at room temperature becomes more of $1/f^{\,2}$ – type at elevated temperatures. The increased frequency power factor $\gamma$ suggests the onset of the electromigration processes. The inset shows the temperature dependent resistance of the quasi-

1D TaSe$_3$ nanowire measured in the range from 300 K to 450 K. The graduate increase of the resistance with temperature, for $T < 410$ K is typical for metal. The sharply rising resistance for T > 410 K indicates the occurrence of electromigration.

**Figure 5:** (a) Extracted frequency power factor $\gamma$ as the function of temperature. The frequency power factor $\gamma$ increases from 1.16-1.24 at 298 K to 1.72-1.75 at 350 K, and remains approximately constant at 1.71-1.75 above 350 K. The blue curve is the fitting of the experimental data. (b) The evolution of the normalized noise spectra density $S_I/I^2$ with temperature $T$. The blue curve is calculated from the Dutta – Horn model. The activation energy estimated from the Dutta - Horn model is $E_P = 1.01$ eV.

**Figure 6:** Temperature dependent $1/f^2$ noise analysis using the Arrhenius plot of $T \times S_I/I^2$ verses $1000/T$. The extracted electromigration activation energy for quasi-1D TaSe$_3$ nanowire is $E_A = 0.88$ eV.

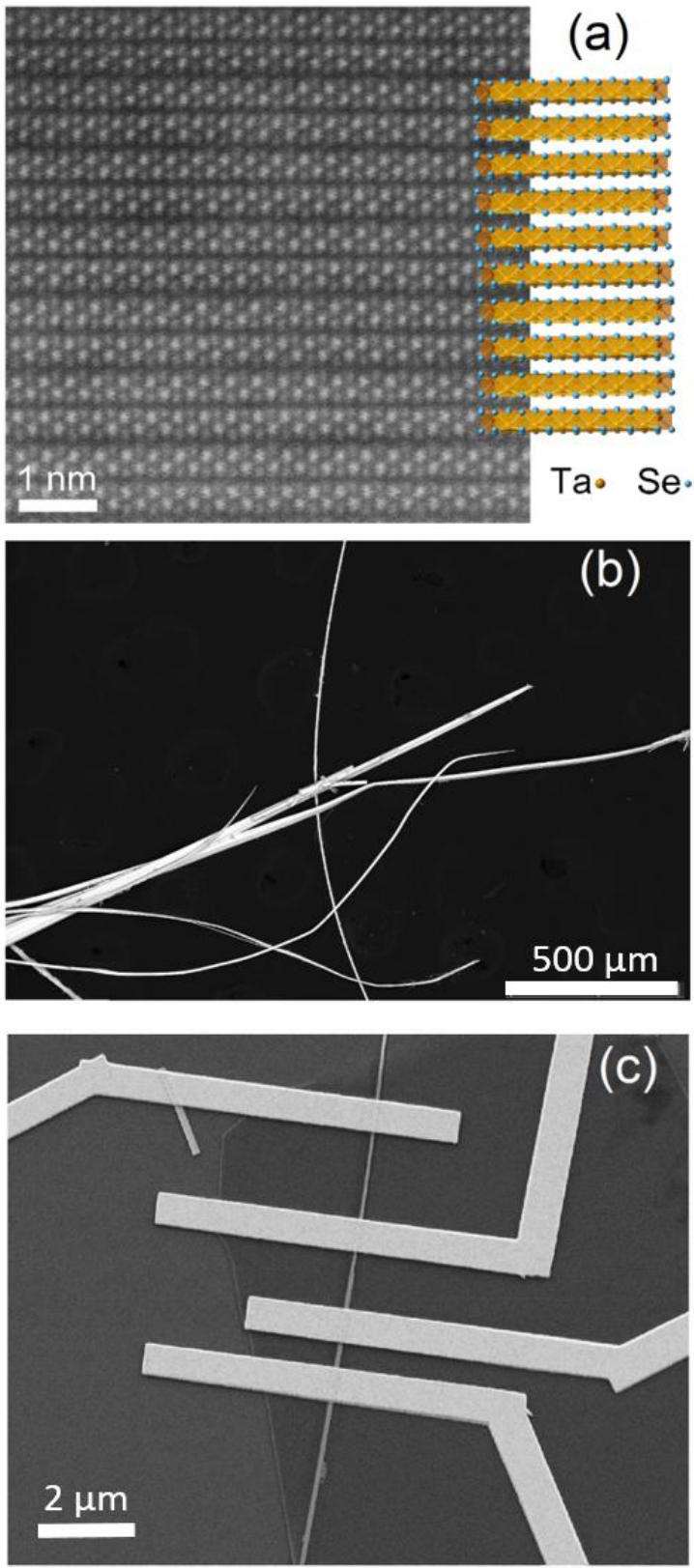

Figure 1

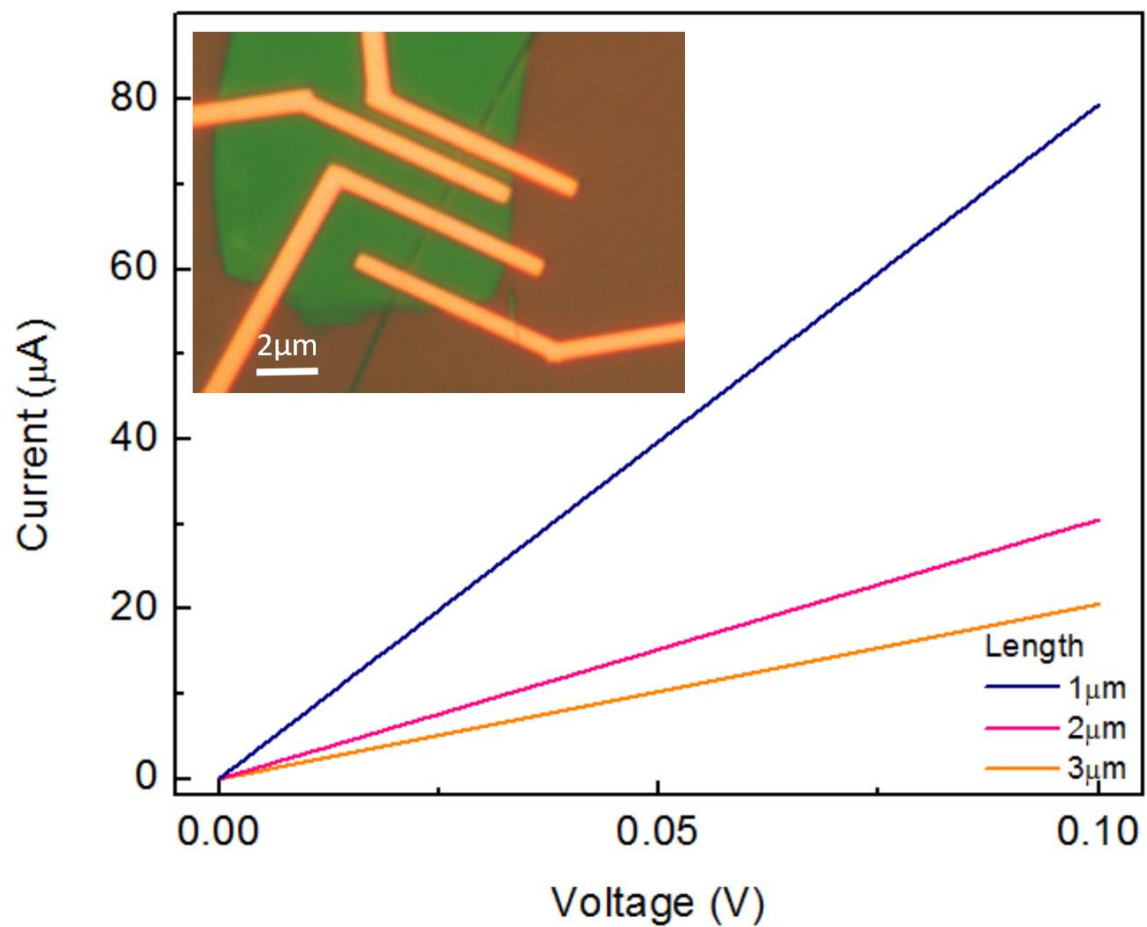

Figure 2

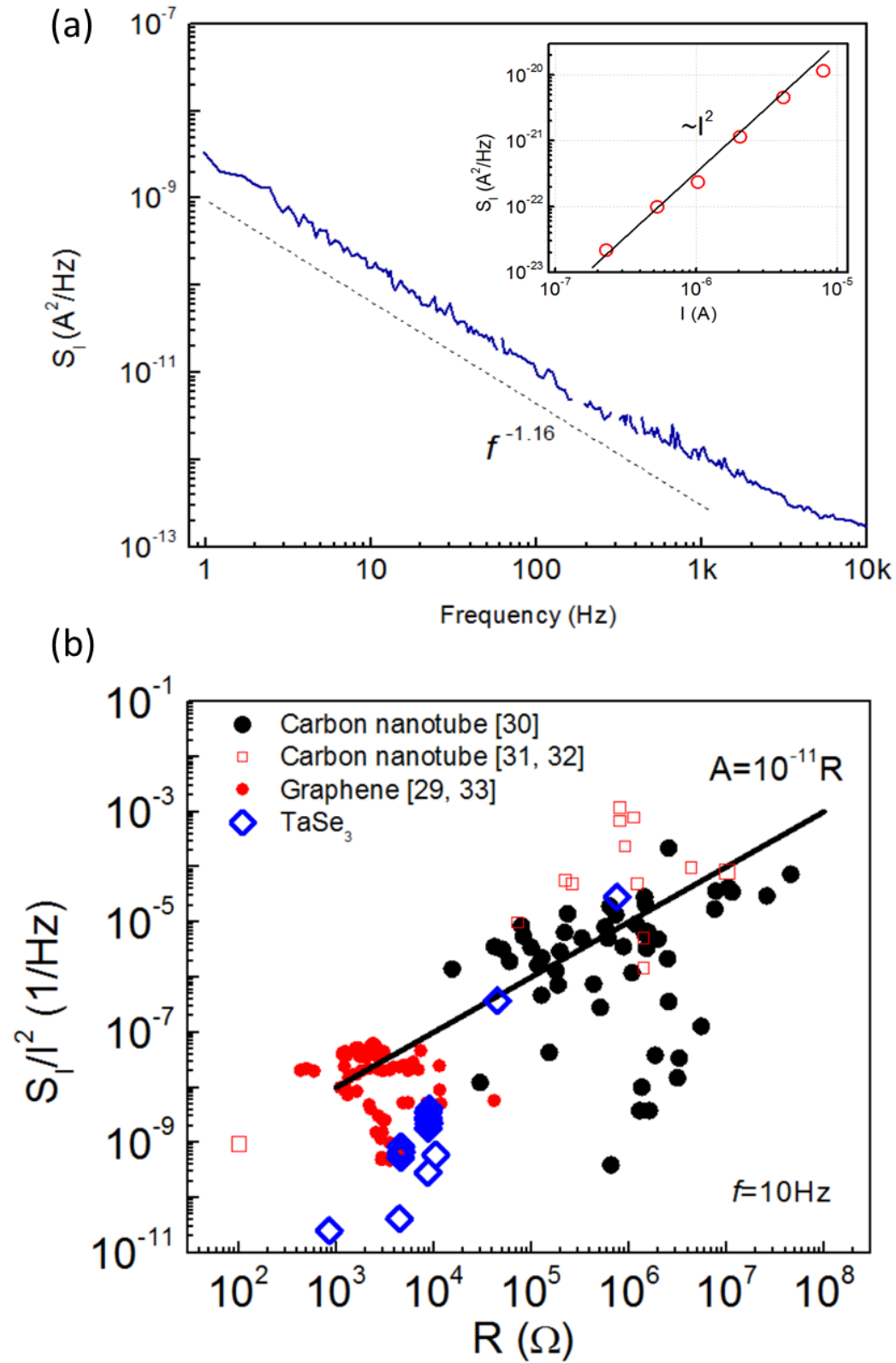

Figure 3

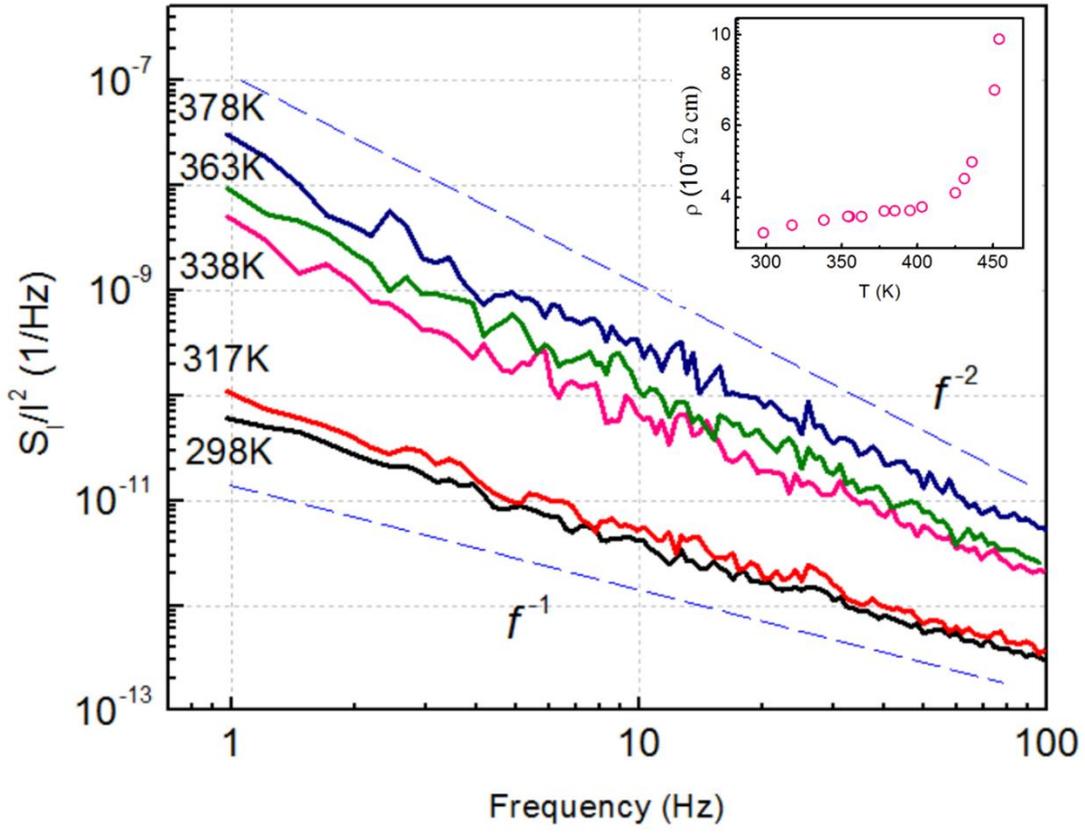

Figure 4

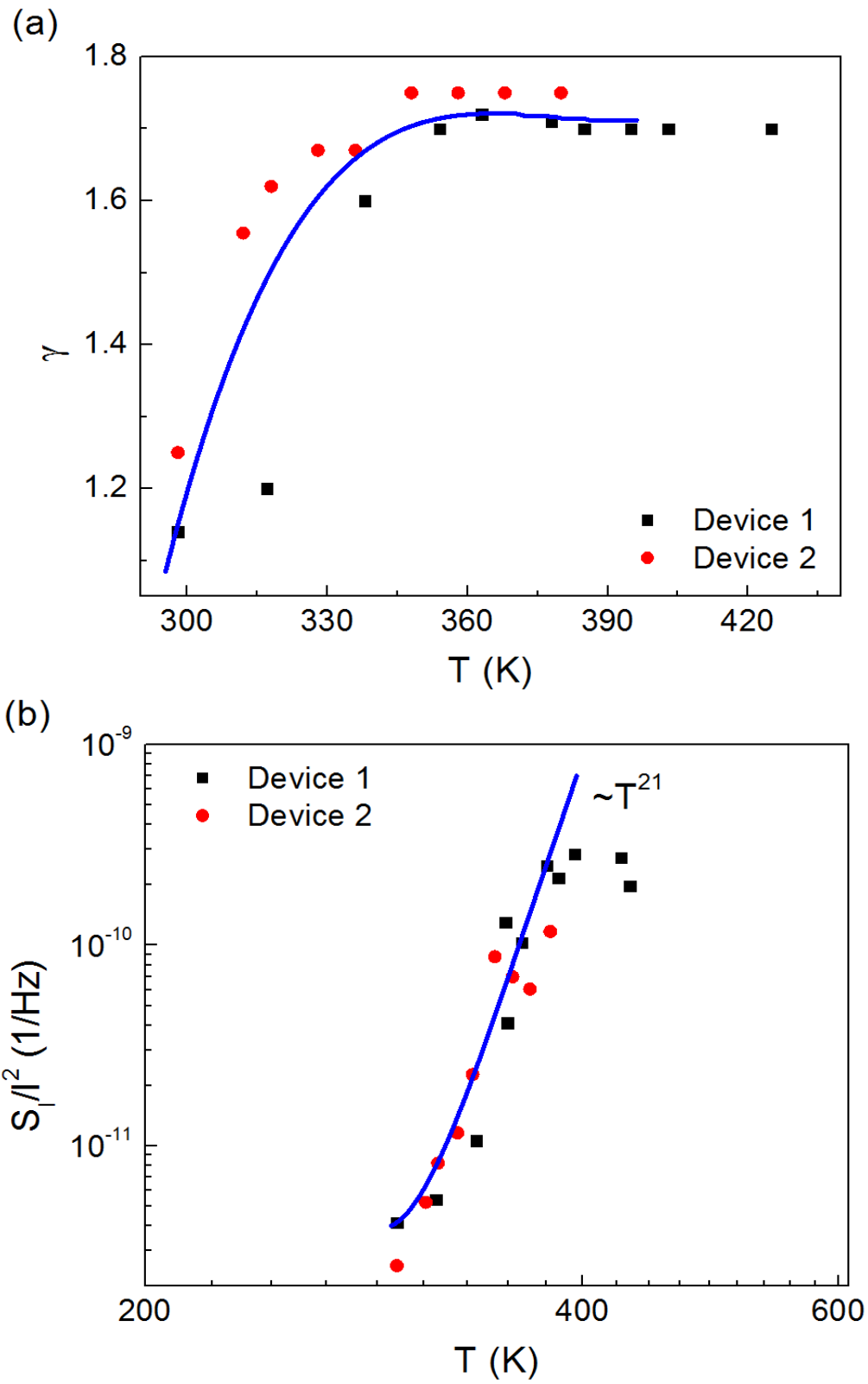

Figure 5

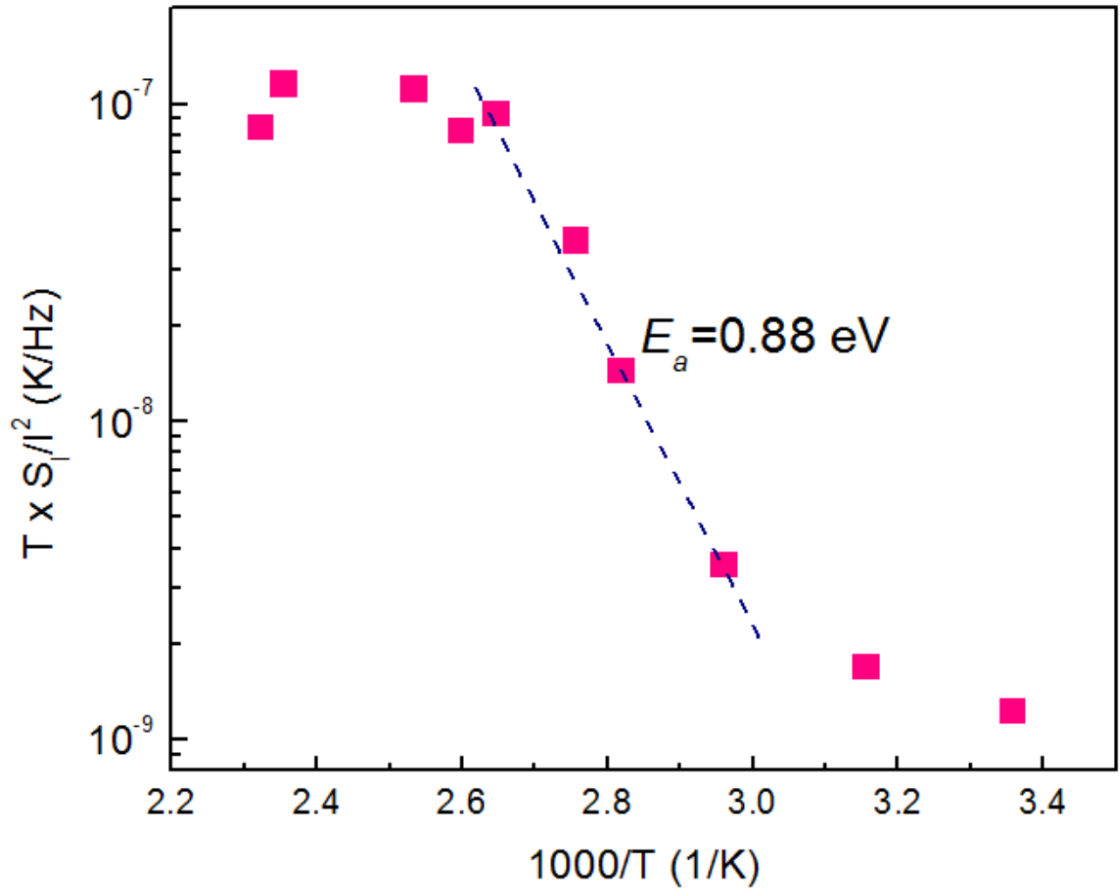

Figure 6